\documentstyle[aps,preprint]{revtex}
\begin{document}
\date{\today}
\preprint{SNUTP 97-050}
\title{Preroughening transitions in  a model \\ 
for Si and Ge (001) type crystal surfaces}
\author{Jae Dong Noh} 
\address{Center for Theoretical Physics, Seoul National University, Seoul
151-742, Korea }
\author{Marcel den Nijs}
\address{Department of Physics, University of Washington, Seattle, WA 98195,
USA}
\maketitle

\begin{abstract}
The uniaxial structure of Si and Ge (001) facets 
leads to nontrivial topological properties of steps and hence to 
interesting equilibrium phase transitions. 
The disordered flat phase and the preroughening transition 
can be stabilized without the need for step-step interactions.
A model describing this is studied numerically by 
transfer matrix type finite-size-scaling of interface free energies. 
Its phase diagram  contains  a flat, rough, and disordered flat phase, 
separated by  roughening and preroughening transition lines. 
Our estimate for the location of the multicritical point 
where the preroughening line merges with the roughening line,
predicts that Si and Ge (001) undergo preroughening 
induced simultaneous deconstruction transitions.

\end{abstract}
\pagebreak

\section{Introduction}\label{sec:1}
The structure of the (001) facets of Si and Ge is very interesting
from the point of view of equilibrium phase transitions. 
These surfaces have an uniaxial reconstruction~\cite{STM}, 
where the uniaxial direction switches by $90^\circ$ at alternating
surface heights. 
Due to this, the mono-atomic and bi-atomic steps have nontrivial 
topological properties. 
This  atomic structure and the crossover from 
mono-atomic steps in non-vicinal surfaces to bi-atomic steps in vicinal 
surfaces have been studied extensively~\cite{Chadi,Aspnes,Alerhand,Pehlke}. 
Roughening type phase transitions in these surfaces
close to the melting temperature, are an  another interesting 
topic~\cite{Johnson,Nijs97}. 
One of us suggested earlier~\cite{Nijs97} that this unusual topology 
leads to disordered flat~(DOF) phases and  preroughening~(PR)
transitions without a need for step-step interactions.
In this paper we present a detailed numerical 
transfer matrix finite-size-scaling~(FSS) study  
of the model introduced in Ref.~\cite{Nijs97}.

Consider a surface like Si (001), but one which does not reconstruct.
Such a surface is still uniaxial and it still switches by $90^\circ$ 
at alternating surface heights. 
At finite temperature $T$, thermodynamically-excited steps appear.
They separate domains of flat regions. 
The uniaxial structure leads to two distinct types of mono-atomic steps, 
labeled by $S_A$ and $S_B$. 
The subscripts denote whether the uniaxial direction in the
upper terrace near the step is parallel~($A$) or normal~($B$) to the step edge. 
Considering the fact that the uniaxial direction switches by $90^\circ$ at
alternating surface heights, one finds, as shown in Fig.~1,
that the steps have the following topological properties ~\cite{Nijs97}: 
(i) If two neighboring parallel steps are of the same type, one must be an
up-step and the other a down-step. 
(ii) If a step turns over $90^\circ$ it must change its type, from $S_A$ 
to $S_B$ and vice versa. 
Bi-atomic steps exist as well ~\cite{Pehlke}, 
but they are probably free-energetically unfavorable
close to the roughening temperature ~\cite{Nijs97}.

These topological properties imply 
that terrace excitations have an ellipsoid shape, 
and that the long axes of nested terraces are parallel (perpendicular)
if the height change is up-down or down-up  (up-up or down-down) across the
nested terraces.
This creates an entropic penalty against forming hills and valleys.
In other words, it opens the possibility to stabilize a DOF phase
without the need for step-step interactions ~\cite{Nijs97}.
Thus far step-step interactions were believed to be crucial for the existence 
of DOF phases. In this surface topology, however,
the DOF phase originates directly from the uniaxial structure of the
surface.
 
The restricted solid-on-solid~(RSOS) model on a square lattice 
with the Hamiltonian
\begin{eqnarray}\label{Hamiltonian}
{\cal H} &=& \sum_{\bf r} \left\{ K(h(x+1,y)-h(x,y))^2 + \Delta \sin\left[
\frac{\pi}{2}(h(x,y)+h(x+1,y))\right]\right\} \nonumber \\
&+& \sum_{\bf r} \left\{ K(h(x,y+1)-h(x,y))^2 - \Delta \sin\left[
\frac{\pi}{2}(h(x,y)+h(x,y+1))\right]\right\} \ ,
\end{eqnarray}
was introduced in Ref.~\cite{Nijs97} 
to describe the thermodynamic properties of such steps in more detail.
$h({\bf r})$ is an integer-valued height variable at each site ${\bf r}=(x,y)$.
Height differences between the nearest 
neighbor sites are restricted to $0$ and $\pm 1$.
This means that only mono-atomic steps are allowed.
Bi-atomic steps  can be included
in a later stage if experimental evidence shows they remain important close 
to roughening temperatures.
The model of Eq.~(\ref{Hamiltonian}) contains two parameters.
The $\Delta$ terms distinguish between $S_A$ and $S_B$-type steps: 
$E_A=K-\Delta$ and $E_B=K+\Delta$ are the step energies.
Without loss of generality, the uniaxial direction is taken to
run vertically~(horizontally) at even~(odd) heights.

The model Hamiltonian contains two limiting cases. 
The conventional RSOS model at $\Delta=0$ displays a Kosterlitz-Thouless~(KT) 
type roughening transition between the flat and rough phases~\cite{Nijs85}. 
On the other hand, in the limit where $E_A=0$ and $T=0$, 
$S_A$ steps cost no energy while $S_B$ steps are frozen out. 
In a typical configuration the surface contains a set of randomly-placed
parallel $S_A$ steps in the form of straight lines. 
However, the topological rule (i)
requires that they are alternating up and down steps. 
This is a typical morphology of surfaces in the DOF phase~\cite{Nijs87}. 
The DOF phase is an intermediate phase between the flat and rough phases, 
where the steps are disordered positionally 
but have long-range up-down-up-down order.
It was argued in Ref.~\cite{Nijs97}
that this DOF type structure  is stable at finite temperatures,
in terms of a Fermionic type perturbation theory. 

In this paper, we investigate the phase diagram quantitatively
through a detailed transfer matrix FSS study. 
It is important to confirm the existence of 
the DOF phase numerically. 
The analysis  in Ref.~\cite{Nijs97} was mostly qualitative. 
The other purpose of this work is to obtain a good estimate 
for the critical value  of the ratio $r\equiv E_B/E_A$, 
below which the DOF phase disappears, see Fig.~2.
In real surfaces this ratio takes specific values. 
For example, observations of step fluctuations 
in STM and LEED experiments yield for Si (001) that 
$r\sim 3$~($E_A = 325K$ and $E_B= 1045K$)~\cite{Bartelt}. 
By comparing this ratio with the critical value $r_c$, 
one can decide which path the Si (001) surface follows.
 
Our model, Eq.~(\ref{Hamiltonian}), does not incorporate the $2\times 1$ 
type dimerized surface reconstruction of Si and Ge (001).
Therefore it does not describe the competition between surface
reconstruction and surface roughening in those surfaces.
This issue was addressed in Ref.~\cite{Nijs97}.
The preroughening line in Fig.~2 is most likely replaced by
a PR induced simultaneous deconstruction transition and the 
roughening line segment  at $r<r_c$ by a roughening induced
simultaneous deconstruction transition.
A  proper quantitative description of this requires at least
a RSOS model coupled to an Ising model.
We did not study such a model, since the
number of degrees of freedom becomes too large to obtain meaningful
transfer matrix FSS results.
The precise location of $r_c$ in Fig.~2 is the result 
of a delicate entropy balancing act of nested terraces
associated with the peculiar $90^\circ$ switching in the uniaxial direction.
Our value of $r_c$ should be meaningful for Si and Ge (001)
if the coupling with the Ising degrees of freedom 
does not change the value of $r_c$ by too much, 
which is a reasonable assumption.

In Sec.~\ref{sec:2}, we introduce various kinds of interface free energies.
They decompose into the free energies of $S_A$ and $S_B$ type steps, 
and show distinct FSS behaviors in the flat, DOF, and rough phases. 
We obtain the phase diagram Fig.~2, by evaluating these interface 
free energies using the transfer matrix method.
The numerical results and a summary are presented in Sec.~\ref{sec:3}.

\section{Interfaces and the transfer matrix formalism}\label{sec:2}

Consider the model given by Eq.~(\ref{Hamiltonian}) on a finite
$N\times M$ lattice with periodic boundary conditions (PBC's), 
$h(x+N,y)=h(x,y)$ and $h(x,y+M)=h(x,y)$.
The ordered flat phase is commensurate with PBC's.
Other (gauge invariant) boundary conditions (BC's) create frustrations, 
and thus impose steps in the surface.
The interface free energy $\eta$ is defined as the excess free energy per unit
length for each type of BC compared to that of PBC's.
Their FSS behaviors are different in the various phases. 
We obtain the structure of the phase transitions
by studying suitable ones. 

The Hamiltonian in Eq.~(\ref{Hamiltonian}) is invariant under the
global transformations 
\begin{eqnarray}
h({\bf r}) & \rightarrow & h({\bf r}) + 2n \label{h+2}\\
h({\bf r}) & \rightarrow & -h({\bf r})+1~({\rm mod}~ 2) \ , \label{-h+1}
\end{eqnarray}
for all integers $n$. 
So it is natural to consider the following boundary conditions:
step type  BC's with  
$h(x+N,y) = h(x,y)+2$ and $h(x,y+M)=h(x,y)$;
and anti-periodic type BC's with
$h(x+N,y)= -h(x,y)+1~({\rm mod}~2)$ and $h(x,y+M)=h(x,y)$.
We will refer to them as  H1 and H2  respectively.
Similarly, V1 and V2 refer to the same BC's 
but with the roles of the $M$ and $N$ interchanged.
The interface free energies are defined by
\begin{eqnarray*}
\eta_\alpha &=& -\frac{1}{M} \ln \frac{Z_{\alpha}}{Z_{PBC}} 
\ , \ (\alpha = {\rm H1,\  H2})\\
\eta_\beta &=& -\frac{1}{N} \ln \frac{Z_{\beta}}{Z_{PBC}} \ , \ 
(\beta = {\rm V1,\  V2}) \ .
\end{eqnarray*}
with $Z_{\alpha}$ the partition function satisfying the
boundary condition $\alpha$,  and 
all energies and free energies measured in units of $k_BT$. 

Figure~3 shows the topological frustrations induced by these BC's.
H1 and V1  require at least two parallel steps; 
one is an $S_A$ type step and the other an  $S_B$ type step
(see Figs.~3 (a) and (c)).
Therefore, $\eta_{\rm H1}$ and $\eta_{\rm V1}$ decompose into
$\eta_{A} + \eta_{B}$;
with $\eta_{A}$ and $\eta_{B}$ 
the $S_A$ and $S_B$ step free energies.
On the other hand, H2 and V2  can be satisfied 
by configurations with only one $S_A$ type step
(see Figs.~3 (b) and (d)). 
Therefore, $\eta_{\rm H2}$ and $\eta_{\rm V2}$  are equal into  $\eta_{A}$.
 
These interface free energies must behave in a specific way in each
type of phase.
The step free energy $\eta_{B}$ is finite in the flat phase and also in the 
DOF phase, but vanishes in the rough phase. 
The step free energy $\eta_{A}$, is finite in the flat phase, 
but vanishes in both the DOF and rough phase.
Therefore,
in the flat phase,  all  four $\eta_{\alpha}$ ($\alpha$=H1, H2, V1, V2) are 
finite. 
In the DOF phase, 
$\eta_{\rm H1}$ and $\eta_{\rm V1}$ remain finite, but
$\eta_{\rm H2}$ and $\eta_{\rm V2}$ vanish (exponentially with system size).
In the rough phase, all four $\eta_{\alpha}$  vanish 
(as a powerlaw in the infinite $M$ and/or $N$ limit).

The rough phase is a critical phase where its critical fluctuations are
described by the Gaussian model. 
The height-difference correlation
function diverges logarithmically with distance:
$$
\langle (h({\bf r})-h({\bf r'}))^2 \rangle \simeq \frac{1}{\pi K_g} \ln | {\bf
r}-{\bf r'} | \ ,
$$
where $K_g$ is the coupling constant of the Gaussian
model (also called the stiffness constant). 
It varies continuously in the rough phase and takes the universal value 
$\frac{\pi}{2}$ at roughening transitions. 
The interface free energies vanish in the rough phase as powerlaws. 
In a semi-infinite geometry $M\rightarrow\infty$, 
$\eta_{\rm H1}$ and $\eta_{\rm H2}$ scale asymptotically as~\cite{Nijs87}
\begin{eqnarray}
\eta_{\rm H1} &=& \frac{2\zeta K_g }{N} \label{Reta_h1} \\
\eta_{\rm H2} &=& \frac{\pi \zeta}{4N} \label{Reta_h2} \ ,
\end{eqnarray}
where $\zeta$ is the aspect ratio of the lattice constants in the
spatial and time-like directions.

We evaluate the interface free energies through the transfer matrix. 
Consider the transfer matrix for a square lattice rotated
by $45^\circ$ as shown in Fig.~4.
In our units the aspect ratio  is equal to $\zeta=2$;
one unit in time, $a_\tau$ is twice as big as the spatial unit $a_x$.
A height configuration $(h_0,h_1,\ldots,h_N)$ in 
a row is represented by a state vector $|h_0,h_1,\ldots,h_N\rangle$. 
It is convenient to replace the height variables by 
step variables $s_i\equiv h_i-h_{i-1}$, with $i=1,\ldots,N$. 
They take only the values $0$, and $\pm 1$ due to the restricted 
solid-on-solid condition.
The surface configuration in each  row  is  therefore  represented by
$|h_0,{\bf s}\rangle$ where ${\bf s}$ stands for $(s_1,s_2,\ldots,s_N)$. 
The elements of the transfer matrix ${\bf T}$ 
are the Boltzmann weights associated with height configurations
$|h_0,{\bf s}\rangle$ and $|h_0',{\bf t}\rangle$ in successive rows.
${\bf T}$ is sparse, and can  be expressed
in terms of a product over local vertex-type scattering matrices, acting on the
$|h_0,{\bf s}\rangle$ and $|h_0',{\bf t}\rangle$ in successive rows 
and intermediate internal step variables ${\bf u}$, defined in Fig.~4. 

In the case of PBC's the step variables satisfy  the conditions
$s_{i+N}=s_i$ and $\sum_{i=1}^N s_i =0$  in all rows. 
The partition function, $Z_{\rm PBC}={\rm Tr}~{\bf T}_{\rm PBC}^M$,
and the free energy, are obtained from the largest eigenvalue 
$e^{-E_{\rm PBC}}$ of ${\bf T}_{\rm PBC}$
in the $M\rightarrow\infty$ limit.

The transfer matrices ${\bf T}_{\rm H1}$ and ${\bf T}_{\rm H2}$ for 
the horizontal BC's H1 and H2 are easily defined. 
Only the conditions the step variables must satisfy change:
In the case of H1, the step variables are again periodic, $s_{i+N}=s_i$,  
but with $\sum_{i=1}^Ns_i= 2$ in all rows. 
In the case of H2, the step variables are anti-periodic, $s_{i+N}=-s_i$,
with no restriction in the value of  $\sum_{i=1}^Ns_i$.
The partition function in each case is given by 
$Z_{\alpha} = {\rm Tr}~{\bf T}_{\alpha}^M$ and the free energy, in the
$M\rightarrow\infty$ limit, is again obtained from the largest eigenvalue
$e^{-E_\alpha}$ of ${\bf T}_\alpha$~($\alpha=$ H1 and H2). 
So the interface free energies are given by
\begin{eqnarray}
\eta_{\rm H1} &=& E_{\rm H1} - E_{\rm PBC} \label{eta_h1}\\
\eta_{\rm H2} &=& E_{\rm H2} - E_{\rm PBC} \ . \label{eta_h2}
\end{eqnarray}

The transfer matrices for the two vertical BC's are more intricate.
They involve the symmetry properties Eq.~(\ref{h+2}) and  Eq.~(\ref{-h+1})
of the transfer matrix with PBC's.
The translation invariance in the surface heights,  Eq.~(\ref{h+2}), 
implies that ${\bf T}_{\rm PBC}$  commutes with the symmetry operator 
\begin{equation}
{\bf P} | h,{\bf s}\rangle = |h+2,{\bf s}\rangle \ .
\end{equation}
Therefore, it is useful to  distinguish between two classes of surface states,
$\{|e,{\bf s}\rangle\}$ and $\{|o,{\bf s}\rangle\}$, 
i.e., all states with $h$ even and odd, respectively.
From the parity type symmetry property  Eq.~(\ref{-h+1})
it follows that  ${\bf T}_{\rm PBC}$ 
commutes also with the operator {\bf R}, defined by
\begin{eqnarray*}
{\bf R} |e,{\bf s}\rangle &=& |o,-{\bf s}\rangle \\
{\bf R} |o,{\bf s}\rangle &=& |e,-{\bf s}\rangle 
\end{eqnarray*}
where $-{\bf s}$ stands for $(-s_1,-s_2,\ldots,-s_N)$. 
The transfer matrices for the vertical BC's can be expressed in term of
${\bf T}_{\rm PBC}$, ${\bf P}$, and ${\bf R}$ as
$ Z_{\rm V1} = {\rm Tr}~[{\bf T}_{\rm PBC}^M {\bf P}]$ and 
$ Z_{\rm V2} = {\rm Tr}~[{\bf T}_{\rm PBC}^M {\bf R}]$.

To evaluate $Z_{\rm V1}$ one needs to  keep track of the height in the first 
column modulo 4.
This makes this boundary condition less useful than its
horizontal counter part  H1, 
where we do not need to keep track of the
absolute height of the surface, and therefore can drop the $h_0$ label 
of the surface configurations altogether.
So we discard $Z_{\rm V1}$ in the following analysis. 

On the other hand, $Z_{\rm V2}$ is very useful.
It can be written as
\begin{equation}
Z_{\rm V2} = \sum_i e^{-ME_{\rm PBC}(i)} - \sum_i e^{-ME_{\rm PBC}'(i)} \ ,
\end{equation}
where $e^{-E_{\rm PBC}(i)}$~($e^{-E_{\rm PBC}'(i)}$) is the $i$th largest 
eigenvalue of ${\bf T}_{\rm PBC}$ in the $R=+1~(-1)$ sector. By the
$R=+1~(-1)$ sector, we mean the set of state vectors which are eigenvectors 
of {\bf R} with the eigenvalue $+1~(-1)$.
Unlike  horizontal boundary conditions, $\eta_{\rm V2}$ 
depends on the entire eigenvalue spectra. 
However, in the thermodynamic limit,
it can be approximated, up to the leading order, as 
$$
\eta_{\rm V2} \simeq -\frac{1}{N} \ln 
\frac{ e^{-ME_{\rm PBC}}-e^{-ME_{\rm PBC}'}}
{e^{-ME_{\rm PBC}}+e^{-ME_{\rm PBC}'}} \ ,
$$
where $E_{\rm PBC}$~($E_{\rm PBC}'$) is the largest eigenvalue of 
${\bf T}_{\rm PBC}$ in the $R=+1~(-1)$ sector.
So the scaling behavior of $\eta_{\rm V2}$ is determined from the quantity
\begin{equation}\label{mgap}
m = E_{\rm PBC}'-E_{\rm PBC} \ ,
\end{equation}
i.e., the mass gap between the two $R$-sectors.
From the fact that $\eta_{\rm V2}$ is finite in
the flat phase and vanishes in the DOF and rough phases, it follows that
this mass gap should be finite 
in the DOF phase
and vanish in the flat phase. 
We will use both   V2 and H2  to locate the 
preroughening phase boundary.

\section{Numerical Results and Summary}\label{sec:3}

The largest eigenvalues in each sector of the transfer matrix
are obtained using the conventional iteration method. 
An arbitrary initial vector projects onto the largest eigenvector
by applying the transfer matrix repeatedly.
$E_{\rm PBC}$, $E_{\rm H1}$, and $E_{\rm H2}$ are easily found by this method.
$E_{\rm PBC}'$ is obtained by choosing the initial vector 
in the $R=-1$ sector.
The state vector is $(2\times 3^N)$ dimensional for an
semi-infinite strip of width $N$.  
The maximum strip width we can handle is $ N=12$.

First, we focus on particular paths through the parameter space  to
illustrate the existence of the rough, DOF, and the flat phases. 
The FSS amplitude 
\begin{equation}\label{s_h1}
S_{\rm H1}(N) \equiv \frac{\eta_{\rm H1}N}{2\zeta} \ 
\end{equation}
of the H1 type interface must  converge to $K_g$ in the rough 
phase~(see Eq.~(\ref{Reta_h1})).
Numerical data of $S_{\rm H1}(N)$ along the line $E_A=0.1$ are presented in 
Fig.~5 (a). It shows that $\eta_{\rm H1}$ scales as $\frac{1}{N}$
with continuously-varying amplitudes at small $E_B$ (high temperatures). 
The $\frac{1}{N}$ scaling breaks down  at large $E_B$. 
This means that $\eta_A$ or $\eta_B$ becomes nonzero.
The roughening transition should take place when 
$S_{\rm H1}(N)$ reaches the universal Kosterlitz-Thouless
value  $\frac{\pi}{2}$.
This value is marked in Fig.~5 (a) by a dashed line,
and indeed it crosses the numerical curve  in the crossover region.
So the numerical data in Fig.~5 (a) support that there is the rough phase 
at high temperatures, 
separated from the DOF or flat phase through a  KT roughening transition. 

We present also the FSS amplitudes of the interface free energy 
$\eta_{\rm H2}$ and the mass gap $m$, defined by
\begin{eqnarray}
S_{\rm H2}(N) &\equiv& \frac{\eta_{\rm H2}N}{\zeta} \label{S_h2} \\
x(N) &\equiv& \frac{mN}{\zeta} \ , \label{x(01)}
\end{eqnarray}
along the line $E_B=3.0$ in Figs.~5 (b) and (c). 
Both quantities show  crossing points.
They signal the crossover between two regions.
One where
$\eta_A$, the $S_A$-step free energy, vanishes (at small $E_B$ ) and
one where it is finite (at large $E_B$). 
This confirms the existence of the DOF phase and the PR transition,
since these free energy gaps must be finite in the flat phase 
but converge exponentially to zero in the DOF phase. 
For some reason, the convergence for V2 is dramatically better than for H2.

At KT type  roughening transitions, the stiffness constant takes the universal
value  $\frac{\pi}{2}$. 
So we obtain a sequence of estimates of the roughening transition line 
by applying the condition $S_{\rm H1}(N) = \frac{\pi}{2}$ for each $N$.
These are the roughening data points shown in  Fig.~2. 
A sequence of estimates for the  PR transition line between 
the flat and DOF phase can be obtained
from the crossing points of $x(N)$ and $x(N-2)$,
and also those of $S_{\rm H2}(N)$ and $S_{\rm H2}(N-2)$.
In Fig.~2  we show only the crossing points of  V2 for $N=6,8,10$ and $12$.
(Those of H2 are much less convergent,  see Fig.~5 (b) and (c)).

The scaling theory of  PR transitions,
tells us that the critical fluctuations are 
described by the Gaussian model, but with $K_g$ greater than the 
universal KT value $\frac{\pi}{2}$ of the roughening transition~\cite{Nijs91}. 
We investigate this scaling behavior
by studying the FSS amplitudes of $\eta_{\rm H2}$ and $m$.
In the Gaussian model, $S_{\rm H2}(N)$, does not vary continuously, 
instead it takes the universal value $\pi/4$ (see Eq.~(\ref{Reta_h2})). 
On the other hand,
the V2 type mass gap should scale as~\cite{unp}
\begin{equation}\label{m_half}
m = \frac{\pi^2\zeta}{2K_gN} \ .
\end{equation}
This is related to the fact that 
at $\Delta=0$, the $R=-1$ and $R=+1$ sectors of ${\bf T}_{\rm PBC}$
are equivalent apart from a phase factor $e^{i\pi s_1}$ 
attached to all step variables $s_1$ at the seam (the first column of the 
lattice). 

In Fig.~6 (a) we present the FSS amplitude of the mass gap $m$.
The vertical axis is scaled as $\pi^2\zeta/(2mN)$, 
such that it represents  $K_g$, see Eq.~(\ref{m_half}). 
$K_g$ starts-off close to the universal value $\frac{\pi}{2}$ 
in the neighbourhood of the roughening transition, at $E_B\simeq 0.8$, 
and increases with $E_B$. 
This is in accordance with the assertion that the 
PR transition is described by the Gaussian model with continuously 
varying $K_g$ greater than $\frac{\pi}{2}$.
The FSS behaviour of $S_{\rm H2}$ along the PR line
is shown in Fig.~6 (b).  The dashed line denotes the universal value 
$\frac{\pi}{4}$ of the Gaussian model. 
The data at large $E_B$ remain far from the universal value, though approach it.
Like before, the convergence of this quantity  is poor (see also Fig.~5(b)). 

The PR transition line in Fig.~2 seems to penetrate into the rough phase. 
But this does not mean that there is another transition inside the rough phase. 
The rough phase is a critical phase where the mass gap scales as 
$O(1/N)$ on either side of the  crossing  points.
The presence of crossing points of V2 inside the rough phase represents only a 
turn around in the corrections to scaling amplitudes for the amplitude.

The crossing of the two sets of lines in Fig.~2,
the estimates for the roughening and PR lines,
confirms the existence of a multicritical point
$(E_{A_c},E_{B_c})$ where the PR and roughening transition merge.
A sequence of estimates $(E_{A_c}(N),E_{B_C}(N))$ for the
multicritical point is obtained by solving the two conditions 
$S_{\rm H1}(N)=\frac{\pi}{2}$ and $x(N) = x(N-2)$ simultaneously
for each value of $N=6,8,10$, and $12$.
These estimates are shown in Fig.~7.
The arrows point towards power-law extrapolated values:   
\begin{eqnarray}
E_{A_c} &=& 0.41\pm 0.03 \\
E_{B_c} &=& 0.89\pm 0.01
\end{eqnarray} 

In summary, we have investigated the phase transitions in a model system for
Si or Ge (001) type crystal surfaces  with an uniaxial structure that switches
direction at each mono-atomic step.
We obtained the phase diagram from a numerical FSS study of the transfer matrix 
spectra. 
It  consists of flat, rough, and DOF phases.
The unusual topological properties of the surface stabilize the DOF phase 
in the absence of step-step interactions, which are crucial for the
stabilization of the DOF phase in conventional surfaces.
The location of the multicritical point where the PR transition line merges 
to the roughening transition line is
determined numerically, $r_c \simeq 2.2$.  

Specific crystals follow paths through Fig.~2 resembling straight lines 
as function of temperature,
since  the step energies are approximately constant.
Our results shows that if the ratio $r=E_B/E_A$ is
greater than a critical value $r_c \simeq 2.2$, 
the flat unreconstructed crystal undergoes a
PR transition into the DOF phase followed by a roughening transition. 
In Si (001) surfaces, the ratio between the
step energies is larger, $r\simeq 3$ ~\cite{Bartelt}.  
We suggest therefore that Si (001) undergoes PR induced simultaneous
deconstruction transition~\cite{Nijs97}. 

\section*{Acknowledgments}
We like to thank Prof.~Doochul Kim for many helpful discussions and his 
hospitality during MdN's visit to the CTP at Seoul National University.
This research is supported by NSF grant DMR-9700430, 
and by Korea Science and Engeneering Foundation through the Center for 
Theoretical Physics, Seoul National University.

\begin{figure}\label{fig:1}
\caption{Topology of $S_A$ and $S_B$ type step 
excitations on an unreconstructed Si (001) type surface.}
\end{figure}

\begin{figure}\label{fig:2}
\caption{ 
The phase diagram of the model defined in Eq.~(1). 
The roughening transition lines
are obtained from $S_{\rm H1}=\frac{\pi}{2}$ for $N=6~(\Box), 8~(\circ)$, and
$10~(\bigtriangleup)$. Preroughening transition lines are obtained from the 
crossing points of $x(N)$ and $x(N-2)$ for $N=6~( \Box),
8~(\circ), 10~(\bigtriangleup)$, and $12~(\bigtriangledown)$. 
The lines between the data points are guides to the eyes.}
\end{figure}

\begin{figure}\label{fig:3}
\caption{
Step excitation type frustrations 
induced by the H1 (a), H2 (b), V1 (c), and V2 (d) boundary conditions. 
The uniaxial direction in each domain of flat region is
shown to help identifying the steps.}
\end{figure}

\begin{figure}\label{fig:4}
\caption{
The transfer matrix set-up.
The rows of a square lattice are rotated by $45^\circ$. 
The aspect ratio $\zeta$ between the lattice constants in the horizontal 
and vertical directions is equal to 2. 
Height and step variables are defined on faces and bonds of the lattice, 
respectively.}
\end{figure}

\begin{figure}\label{fig:5}
\caption{
Typical data of the FSS amplitude of the interface free energies 
$\eta_{\rm H1}$ (a), $\eta_{\rm H2}$ (b) and the mass gap $m$ (c).
Different symbols~($\Box$ for $N=6$, $\circ$ for $N=8$, $\bigtriangleup$ for 
$N=10$, and $\bigtriangledown$ for $N=12$) are used to 
distinguish the strip widths $N$. 
The lines are guides to the eyes.}
\end{figure}

\begin{figure}\label{fig:6}
\caption{
FSS amplitudes of $m$ (a) and $\eta_{\rm H2}$ (b) 
along the preroughening transition line. Different symbols~($\Box$ for
$N=6$, $\circ$ for $N=8$, $\bigtriangleup$ for
$N=10$, and $\bigtriangledown$ for $N=12$) are used to
distinguish the strip widths $N$.}
\end{figure}

\begin{figure}\label{fig:7}
\caption{
Estimates $(E_{A_c}(N),E_{B_c}(N))$ for the location of the 
multicritical point for $N=6,8,10$, and $12$. 
The extrapolated values are marked by arrows.}
\end{figure}

\end{document}